\documentclass[prl,twocolumn,showpacs,amsmath,amssymb]{revtex4}
\usepackage{graphicx}
\usepackage{dcolumn}
\usepackage{bm}

\begin{document}

\title{Laser Ion Acceleration Scaling Laws Seen in Muti-Parametric PIC Simulations}

\author{T. Esirkepov}
\author{M. Yamagiwa}
\author{T. Tajima}
\affiliation{Kansai Research Establishment, JAEA, Kizu, Kyoto, 619-0215 Japan}

\date{October, 2005}

\begin{abstract}
The ion acceleration
driven by a laser pulse at intensity
$I = 10^{20} - 10^{22}\;$W/cm$^2\times(\mu$m$/\lambda)^2$
from a double layer target
is investigated
with multi-parametric Particle-in-Cell (PIC) simulations.
For targets with a wide range of thickness $l$ and density $n_e$,
at given intensity 
the highest ion energy gain
occurs at certain {\it critical depth} of the target  $\sigma = n_e l$,
which is proportional
to the square root of intensity.
In the case of thin targets and optimal laser pulse duration,
the ion maximum energy scales as the
square root of the laser pulse power.
When the radiation pressure of the laser field becomes dominant,
the ion maximum energy becomes proportional to the laser pulse energy.
\end{abstract}



\maketitle


The laser-driven ion acceleration
can be of benefit for many applications,
e.~g.,
hadron therapy in oncology \cite{Bulanov-Khoroshkov},
fast ignition of thermonuclear fusion by protons \cite{Roth},
production of sources for positron emitting tomography (PET) \cite{PET},
conversion of radioactive waste \cite{Nucl-waste},
etc.
The radiation pressure dominant (RPD) regime
of the laser ion acceleration \cite{Cocoon}
can be a basis for a
laser-driven heavy ion collider
and proton dump facility for neutrino oscillation studies \cite{Neutrino}.
All these applications can become possible owing to
invention of high-intense lasers \cite{Mourou},
that now are capable to produce
pulses with intensity $10^{20}-10^{22}\,$W/cm$^2$.

Irradiation of solid targets
by lasers with power from terawatt (TW) to petawatt (PW)
and intensity up to $10^{21}\,$W/cm$^2$
resulted in generation of proton beams with
maximum energy ${\cal E}_{\rm max}$ from 1 MeV up to 55 MeV \cite{PW-proton}.
Typical proton beam has
picosecond duration
and is emitted in quasilaminar fashion \cite{Borghesi};
it can have very low transverse emittance \cite{Cowan}.
Experiments indicate that
the proton energy increases
with the laser intensity 
and it depends also on the target (foil) thickness
and the target composition.
It turns out that for given laser intensity
there is an optimal target thickness,
at which the proton energy is maximum \cite{opt-thick}.
Thus the important question,
what is the ion maximum achievable energy at given intensity
and how it scales with increasing intensity?

In this Letter we investigate
the scaling laws of the ion acceleration driven by
a laser pulse at intensity $I = 10^{20} - 10^{22}\;($W/cm$^2)(\mu$m$/\lambda)^2$
from a {\it double layer target}
with {\it multi-parametric} Particle-in-Cell (PIC) simulations.
The double layer target
was suggested in Refs. \cite{Bulanov-Khoroshkov}
for a laser-driven high-quality ion beam generation
that can be used in hadron therapy;
this scheme was also demonstrated with
three-dimensional PIC simulations in Refs. \cite{DLT}.

In publications on the multi-MeV ion acceleration
driven by an intense laser from foil targets,
several regimes of acceleration can be found.
For example,
ions can be accelerated by varying electrostatic potential
of the electron cloud expanding into vacuum,
as shown in Refs. \cite{Gurevich} with different models.
In the case of thin foils,
the ion acceleration can occur in a regime,
when the laser pulse quickly sweeps significant part of electrons away,
\cite{Bulanov-Khoroshkov,DLT,SPIG}.
At high intensity, when the radiation pressure of the laser field become dominant,
the plasma is accelerated
so that almost all the energy of the laser pulse
is transformed into the ion energy, \cite{Cocoon}.
Below we show that, according to our simulations, 
in the case of thin targets and optimal laser pulse duration,
the ion maximum energy scales as the
square root of the laser pulse power, as suggested in \cite{SPIG}.
With increasing intensity, the RPD regime comes into play
and the ion maximum energy becomes proportional to the laser pulse energy.


The laser-driven ion acceleration also depends
on the structure of the laser pulse.
``Parasitic'' parts of the laser pulse such as
a pre-pulse and an amplified spontaneous emission (ASE)
can substantially change the structure of the solid target before
the main part of the laser pulse arrives,
e.~g., a pre-plasma can be formed.
The effect of ``parasitic'' parts of the laser pulse
on the ion acceleration
and dependence of the ion energy on
the scale-length of the pre-formed plasma
was dicsussed in Refs. \cite{Nemoto, Sentoku, Matsukado}.


We study the simplest model,
where the ``clean'' gaussian p-polarized laser pulse,
whithout ``parasitic'' parts, 
is incident at right angle on a thin slab
of ideal collisionless plasma.
The approximation of collisionless plasma
can be satisfactory when a femtosecond laser
at intensity $\gtrsim 10^{20}\,$W/cm$^2$
interacts with plasma slab,
which is a few laser wavelengths thick,
even at solid density of plasma.
The laser pulse is characterized by
the wavelength $\lambda$,
the dimensionless amplitude $a$,
corresponding to intensity
$I=a^2 I_{1}$,
$I_{1} = 1.368\!\times\! 10^{18}\,$W/cm$^2\!\times\!(\mu$m$/\lambda)^2$,
the length (FWHM) $L$
and focal spot diameter (FWHM) $D$.
The target consits of two layers,
the first layer is fully stripped aluminium
and the second layer is a proton coating.
The first and second layers of the target
are described, respectively, by
the electron density $n_e$ and $n_{e2}$,
the thickness $l$ and $l_2$,
the transverse (perpendicular to the laser pulse direction)
size $\mbox{\sl w}$ and $\mbox{\sl w}_2$.

We carried out two-dimensional (2D) PIC
multi-parametric simulations
using a technique described in \cite{Matsukado}.
In this technique a series of 2D PIC tasks with $N$ different sets of
the laser and target parameters
is performed simultaneously on the $N$ processors of
a multi-processor supercomputer,
using the massively-parallel
and fully vectorized code REMP,
based on the PIC method and
the ``density decomposition'' scheme \cite{CPC2001}.
We analyze the dependence of the interaction outcomes
such as reflection, transmission and absorption coefficients,
maximum ion and electron energy and kinetic temperature,
beam emittance, energy spread,
the acceleration time and length, etc.
on parameters of the laser pulse and the target:
laser pulse intensity $I$, focal spot size $D$ and duration $L/c$,
target density $n_e$ and thickness $l$.
In the simulations,
the laser pulse initially propagates along the $x$-axis
in the simulation box with $x\times y$ size
$251\lambda\times 136\lambda$;
the transverse size of the target first layer is fixed:
$\mbox{\sl w} = 80\lambda$,
for the second layer it is $\mbox{\sl w}_2=D/2$;
the second layer is $0.06\lambda$ thick,
its density is such that the number of ions
in the first layer in the longitudinal direction 
is $\sim 10^3$ times greater than the corresponding number of protons.
At chosen conditions,
protons behave almost like probe particles,
thus we can analyze, almost independently,
both the proton acceleration
and the acceleration of heavier ions from the first layer.
Simulations were performed on 720-processor HP AlphaServer at JAERI-Kansai.
In each processor the grid consists of $4016\times 2176$ cells
and a number of quasi-particles is $1.9\times 10^{7}$,
amounting to $6.3\times 10^9$ of cells and
$1.4\times 10^{10}$ of quasi-particles in total.

\begin{figure*}
\includegraphics{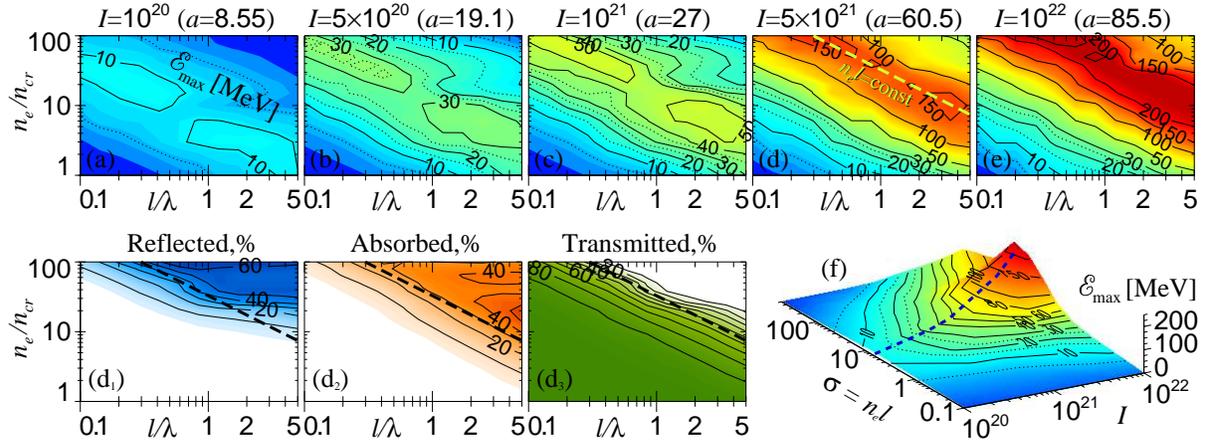}
\caption{\label{fig:E-lna}
Maximum proton energy (contours) vs.
target thickness and density
(log-log scale)
for different laser intensities, (a)-(e),
for $L=10\lambda$, $D=10\lambda$.
The laser pulse reflection (d$_1$), absorption (d$_2$)
and transmission (d$_3$) coefficients,
related to frame (d).
Dashed line in (d), (d$_{1-3}$) -- for the best $n_e l$,
corresponding to maximum energy gain.
(f): maximum proton energy vs laser intensity
and target {\it critical depth} $\sigma=n_e l$ (log-log scale).
Dashed line for optimal $\sigma_{opt}\propto I^{1/2}$.
Intensity unit W/cm$^2\times (\mu$m$/\lambda)^2$.}
\end{figure*}

Fig. \ref{fig:E-lna} shows how the maximum proton energy
depends on the target thickness and density
and the laser intensity for fixed laser length $L=10\lambda$
and focal spot size $D=10\lambda$.
The proton energy was taken at the time when the
proton acceleration dropped to 3\% from maximum;
the corresponding acceleration length is approximately equal
to the laser focal spot size.
Frames (a)-(e) represent results of 720 separate computational tasks
performed simultaneously in one run on 720-processor supercomputer,
each point with coordinates $(l,n_e)$ at fixed intensity $I$
corresponds to particular one-processor task.
The target thickness changes in the range $l/\lambda = 0.1  -  5$ (12 samples),
the target density varies from $n_e = n_{cr}$ to $100 n_{cr}$ (12 samples),
where $n_{cr}=\pi/(r_e \lambda^2)$ is the critical density,
$r_e = e^2/m_e c^2$ is the classical electron radius.
The intensity varies from $I=10^{20}$ to $10^{22}$W/cm$^2\times(\mu$m$/\lambda)^2$
(5 samples).
Plasma slab with comparatively low density and large thickness
can be considered as a model of the target modified by
``parasitic'' parts of the laser pulse.

In fig. \ref{fig:E-lna}
the ion energy curves appears to be nearly parallel to the
contours of the product $\sigma = n_e l$,
i.~e. the number of electrons in the longitudinal direction.
We call this parameter the {\it critical depth} of the plasma slab.
We see that  the energy dependence
on two parameters $l$ and $n_e$
degenerates to a dependence on only one parameter $\sigma$.
Such a degeneracy is rather surprising if we inspect
the results of different simulations with the same $\sigma$.
At small thickness $l$ and large density $n_e$
the laser pulse sweeps away a substantial part of electrons,
the induced strong Coulomb potential of the first layer accelerates protons,
in accordance with the scenario described in Refs. \cite{Bulanov-Khoroshkov,DLT,SPIG}.
At large thickness and small density
the laser pulse penetrates throuhg the target
and generates a strong quasi-static magnetic field
whose pressure causes charge separation which accelerates protons
near the plasma-vacuum interface,
similarly to the mechanism suggested in Refs. \cite{Kuznetsov, Matsukado}.
The maximum energy gain corresponds to some optimum {\it critical depth}
$\sigma_{\rm opt}$.
For decreasing $\sigma < \sigma_{\rm opt}$,
more and more laser pulse energy is transmitted through the plasma slab.
For increasing $\sigma > \sigma_{\rm opt}$,
the laser pulse reflection becomes more and more efficient.
The energy absorption turned out to be optimal for the ion acceleration
at $\sigma=\sigma_{\rm opt}$.
Fig. \ref{fig:E-lna}(f) shows the dependence of the energy gain
on the laser intensity $I$ and the {\it critical depth} $\sigma$.
Varying the $\sigma$,
one can find the minimum intensity which gives
the desired energy gain.
In other words, at each intensity
there is an optimal $\sigma$ which gives the maximum energy gain.
The optimal {\it critical depth} almost linearly depends on the
square root of the laser intensity,
$\sigma_{\rm opt}/n_{cr}\lambda \approx
3+0.4 \times (I/I_{1})^{1/2}$.
This is similar to the criteria of relativistic transparency
of a thin foil, derived in Refs. \cite{rel-trans}.

\begin{figure*}
\includegraphics{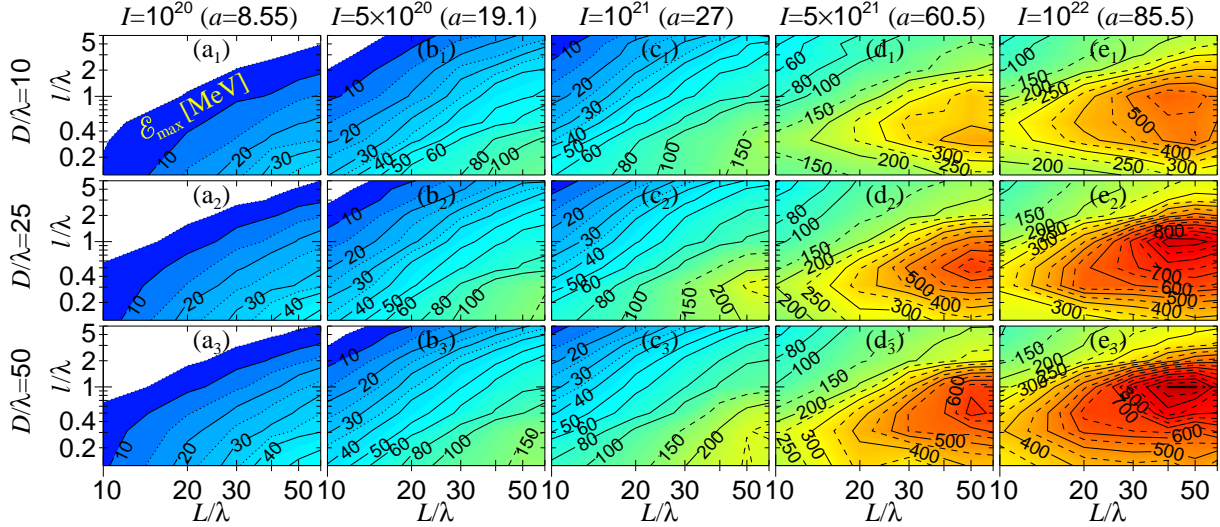}
\caption{\label{fig:E-LlaD}
Maximum proton energy (contours) vs.
laser length and target thickness
(log-log scale)
for different laser intensities (columns (a)-(e))
and different laser focal spots (raws 1-3),
for $n_e = 100n_{cr}$.}
\end{figure*}

Fig. \ref{fig:E-LlaD} shows the maximum proton energy vs the laser length
and the target thickness
for different laser intensities and focal spots
at fixed plasma density $n_e=100 n_{cr}$.
As above, the proton energy corresponds to the time when 
the acceleration damped to 3\% from maximum.
The laser pulse length is chosen in the range $L/\lambda = 10  -  60$ (6 samples),
the target thickness is $l/\lambda = 0.125  -  5$ (8 samples),
the laser intensity samples are the same as above,
the laser focal spot diameter choices are $D/\lambda = 10,25,50$.
We see that the energy gain increases with decreasing target thickness
and increasing laser pulse length.
Columns (b) and (c)
indicate that energy in the range 100-200 MeV
is achievable with a ``clean'' petawatt laser with sub-picosecond
duration when the solid density target is sufficiently thin,
as was found in \cite{MMeV-ions}.

At intensity of the order of
$5\times 10^{21}$W/cm$^2\times (\mu$m$/\lambda)^2$,
the new regime of acceleration comes into play,
when the energy gain rapidly increases with the laser pulse length
and, for optimal pair of laser duration and target thickness,
relativistic protons can be obtained.
This is the {\it radiation pressure dominant} regime of the ion acceleration,
described in \cite{Cocoon} for much higher intensity.
In this case the laser radiation pressure dominates in the interaction
and the effective cross-section of the process
(analog of the Thomson cross-section)
becomes $2/n_e l$, \cite{Cocoon}.
As shown in Fig. \ref{fig:E-LlaD},
the energy gain decreases if the laser duration
is greater than the optimum.
This is because we use Gaussian laser pulses,
so a long pulse head have enough time to deteriorate the target.
As indicated in Refs. \cite{Cocoon},
in the ideal case of the RPD regime,
the energy gain is proportional to the laser pulse duration.

\begin{figure*}
\begin{minipage}{\columnwidth}
\includegraphics{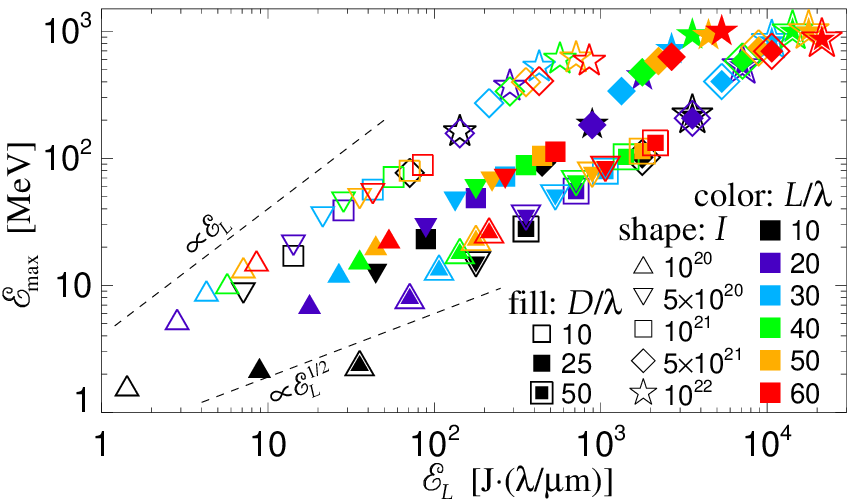}
\caption{\label{fig:E-Elas}
Maximum proton energy vs.
laser pulse energy
for $l = \lambda$, $n_e = 100n_{cr}$.
Dashed lines exemplify possible scalings.}
\end{minipage}
\hfil
\begin{minipage}{\columnwidth}
\includegraphics{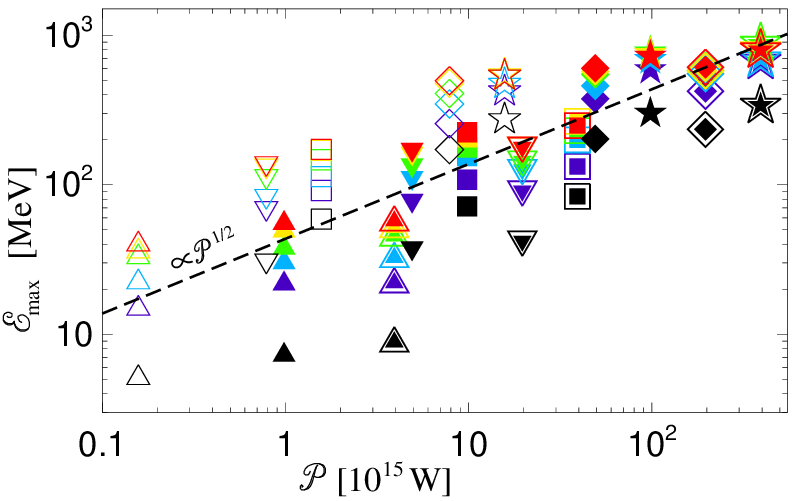}
\caption{\label{fig:E-Pow}
Maximum proton energy vs.
laser power
for optimal plasma slab thickness and $n_e = 100n_{cr}$.}
\end{minipage}
\end{figure*}

Figs. \ref{fig:E-Elas} and \ref{fig:E-Pow}
are compiled from the results presented above.
In Fig. \ref{fig:E-Elas} the maximum proton energy ${\cal E}_{\rm max}$
is drawn for every instance of the laser pulse energy ${\cal E}_{\rm L}$
in the case $l = \lambda$, $n_e = 100n_{cr}$.
Points become arranged along 3 lines
corresponding to 3 versions of the focal spot size;
lines can be fitted by scaling ${\cal E}_{\rm max} \propto {\cal E}_{\rm L}^{0.8}$
up to ${\cal E}_{\rm max}\lesssim 200\,$MeV
and by ${\cal E}_{\rm max} \propto {\cal E}_{\rm L}$
for higher energy gains when begins a transition to the RPD regime.

Fig. \ref{fig:E-Pow} shows a correlation of the
maximum proton energy with the laser power ${\cal P}$
for $n_e = 100n_{cr}$, where
points correspond to a thickness $l$ close to optimal,
i.~e. $n_e l \approx \sigma_{\rm opt}$.
Narrowing the set of points by additional
constraint that the laser pulse length
is of the order of the focal spot size, $L\sim D$,
we obtain a strip aligned along the dependence
${\cal E}_{\rm max} \propto {\cal P}^{1/2}$.
Once again we see a degeneration of a complex dependence
of the ion maximum energy from many parameters
to a dependence from a fewer number of parameters.
This can be a manifestation of a distinct regime.
In this case the conditions are met for the regime
described in Refs. \cite{Bulanov-Khoroshkov,DLT,SPIG}.
The square-root dependence on the laser power
is consistent with the prediction made in Ref. \cite{SPIG}.
Following Refs. \cite{Bulanov-Khoroshkov,DLT,SPIG},
one can represent the irradiated spot as
an uniformly charged oblate ellipsoid
with size $l\!\times\! D\!\times\! D$, $l<D$,
and charge density $+\eta e n_e$, where
$\eta$ is a portion of electrons swept away by the laser pulse.
Assuming that the target {\it critical depth} is optimal,
$n_e l = 0.4\, a\, n_{cr}\lambda$,
one can obtain the maximum energy of the ion with charge $Ze$
accelerated by the electric field of the ellipsoid (for $l\ll D$):
${\cal E}_{\rm max} \approx
\eta Z ({\cal P}[\mbox{PW}])^{1/2}\times 228\,$MeV,
where ${\cal P}$ is the laser peak power.
The condition that the optimal laser pulse length should be
greater than or of the order of the focal spot size, $L\gtrsim D$,
turs out to be necessary
to prevent return currents from affecting the ion acceleration
(both the current, formed by returning electrons,
and that, produced by electrons from a surrounding less irradiated
and relatively cold plasma).


In conclusion,
our multi-parametric simulations show that
in the process of the
ion acceleration by an
intense laser pulse from a double layer target,
at given laser intensity $I$
the dependence of the ion maximum energy
from the target thickness $l$ and density $n_e$
is reduced to a dependence on the {\it critical depth} $\sigma=n_e l$.
At given intensity
the highest ion energy gain
occurs at optimal {\it critical depth} of the target $\sigma_{\rm opt}$
which is approximately proportional
to the square root of intensity, $\sigma_{\rm opt} \propto I^{1/2}$.
If the target {\it critical depth} is less than the optimal value,
the laser pulse is more transmitted rather than absorbed;
if it is greater, the laser pulse reflection increases
making the interaction less efficient for the ion acceleration.
Simulations indicate that
in the case of thin targets and optimal laser pulse duration,
the ion maximum energy scales as the
square root of the laser pulse power,
in consistence with the theory presented in Ref. \cite{SPIG}.
Therefore, the laser-driven ion acceleration up to 200-300 MeV, which
is necessary for hadron therapy, can be obtained with a petawatt laser with
sub-picosecond duration.
A transition to the radiation pressure dominated regime
of the ion acceleration is revealed.
With increasing intensity, when the RPD regime comes into play,
the ion maximum energy becomes proportional to the laser pulse energy.

\begin{acknowledgements}
We thank
Prof. S. V. Bulanov
for the formulation of the problem and helpful suggestions.
We thank
Prof. H. Daido,
Prof. C. Ma,
Prof. G. Mourou,
Prof. Nemoto,
Prof. Noda
for discussions,
MEXT and JST CREST for support.
\end{acknowledgements}


\end{document}